\documentclass[useAMS,usenatbib,18pt]{mn2e}

\usepackage{times}

\usepackage[dvips]{graphicx}

\title[Spatially Resolved Spectroscopy of Passive Spiral Galaxies]{Spatially Resolved Spectroscopy of Passive Spiral Galaxies}
\author[Miho Ishigaki, Tomotsugu Goto and Hideo Matsuhara]{Miho Ishigaki$^{1,2}$\thanks{E-mail: miho@astr.tohoku.ac.jp (M.I.); tomo@ir.isas.jaxa.jp (T.G.); maruma@ir.isas.jaxa.jp (H.M.)}\footnotemark[1]\thanks{Astronomical Institute, Tohoku University, Sendai, 980-8578, Japan},Tomotsugu Goto$^{1}$ and Hideo Matsuhara$^{1,2}$\\
$^{1}$Institute of  Space and Astronautical Science, Japan Aerospace Exploration Agency, 3-1-1 Yoshinodai, \\
Sagamihara, Kanagawa 229-8510, Japan\\
$^{2}$Tokyo Institute of Technology, 2-12-1 Ookayama, Meguro-ku,\\
Tokyo, 152-8550, Japan}

\begin{document}

\maketitle

\label{firstpage}

\begin{abstract}

Passive spiral galaxies, despite their spiral morphological appearance, do not have any emission lines indicative of ongoing star formation in their optical spectra. Previous studies have suggested that passive spiral galaxies preferentially exist in infall regions of galaxy clusters, suggesting that the cluster environment is likely to be responsible for creating these galaxies. By carrying out spatially resolved long-slit spectroscopy on four nearby passive spiral galaxies with the Apache Point Observatory 3.5-m telescope, we investigated the stellar populations of passive spiral galaxies separately for their inner and outer regions. In the two unambiguously passive spiral galaxies among the four observed galaxies, H$\delta$ absorption lines are more prominent in the outer regions of the galaxies, whereas the 4000-{\AA} breaks (D$_{4000}$) are strongest in the inner regions of the galaxies. A comparison with a simple stellar population model for the two passive spiral galaxies indicates that the outer regions of the galaxies tend to harbour younger populations of stars. The strong H$\delta$ absorption observed in the outer regions of the sample galaxies is consistent with that of galaxies whose star formation ceased a few Gyrs ago. Because of the large uncertainty in the absorption indices in our samples, further observations are needed in order to place constraints on the mechanisms that quench star formation in passive spiral galaxies.
\end{abstract}


\section{Introduction}
\label{sec:section1}
Passive spiral galaxies have peculiar spectroscopic characteristics among the galaxy populations having spiral morphologies. They show few or neither of the emission lines in H$\alpha$ nor [O II] in optical spectra that would be indicative of ongoing star formation (Couch et al. 1998; Dressler et al. 1999; Poggianti et al. 1999; Goto et al. 2003b). The optical g $-$ r colours observed in such galaxies are found to be significantly redder than those of spiral galaxies with emission lines, which confirms the lower star formation rate among passive spiral galaxies (Poggianti et al. 1999).

A similar population of galaxies, known as 'anemic' spiral galaxies, has been found in the Virgo Cluster (van den Bergh 1976). They have smoothed spiral arms showing less prominent star formation activity than in other galaxies of the same Hubble type. Elmegreen et al. (2002) observed anemic spirals in the Virgo Cluster and found lower gas surface densities than for normal spirals for these galaxies. They showed that the gas surface density in these galaxies is below the threshold for star formation (Kennicutt 1989), suggesting that the lack of star formation is caused by the stripping of gas in the environment of the cluster (Elmegreen et al. 2002).

In order to investigate the effect of environment on the formation of such galaxies, Goto et al. (2003b) searched for passive spiral galaxies in all environments, including in dense cluster cores and field regions, using the large samples in the Sloan Digital Sky Survey (SDSS) data (Strauss et al. 2002). They found that passive spiral galaxies preferentially exist in the environment where a local galaxy density is intermediate between that of the cluster cores and the field regions (Goto et al. 2003b). This characteristic environment corresponds to the region where a significant decline in the star formation rate has been observed (Lewis et al. 2002; Gomez et al. 2003; Tanaka et al. 2004). These studies imply that cluster-related phenomena could be the main factors responsible for the formation of passive spiral galaxies.

Recent observations have revealed that the cluster environment may bring about a transformation in the galaxy population from star-forming spirals to passive galaxies, and thus passive spiral galaxies have been suggested to be a population in a transitional phase between these two populations (Poggianti et al. 1999). For instance, the fraction of blue galaxies is found to be larger in distant clusters than in local clusters, a phenomenon known as the Butcher-Oemler effect (Butcher \& Oemler 1978; Couch et al. 1998; Kodama \& Bower 2001; Goto et al. 2003d, 2004). The Butcher-Oemler effect implies that the environment of a cluster may affect the star formation activity of its member galaxies. Furthermore, morphological type (Goto et al. 2003c; Treu et al. 2003; Postman et al. 2005) is found to be correlated with various functions of cluster environment, such as local density or cluster centric radius (Goto et al. 2004; Tanaka et al. 2004). Numerical simulations (Bekki et al. 2002) have shown that gas-stripped galaxies may finally become S0 galaxies if no further accretion onto the disc occurs after the stripping. UV observations suggest that the cessation of star formation could take place before this morphological transformation into S0 galaxies (Moran, Ellis \& Treu 2006). Indeed, Goto et al. (2003b) found that the Petrosian radius of S0s is smaller than that of spirals, which is consistent with the above scenario. Although these studies have provided significant broad implications concerning the origin of passive spiral galaxies, the details of the underlying physical mechanisms, especially the time-scale on which in-falling cluster galaxies terminate their star formation, are still uncertain.

Various mechanisms with different time-scales have been proposed. Because star formation could be sustained by cold gas accreted onto the disc, the time-scale is closely related to the time at which the reservoir of cold gas is removed by means of galaxy-intracluster medium (ICM) interactions (Treu et al. 2003; Diaferio et al. 2001; Kauffmann et al. 1999).

A possible mechanism that occurs on a relatively short time-scale is ram-pressure stripping (Gunn \& Gott 1972; Fujita 2001; Fujita \& Nagashima 1999; Fujita \& Goto 2004; Abadi, Moore \& Bower 1999; Quilis, Moore \& Bower 2000). When a galaxy falls into a dense region of the cluster, ram pressure caused by the motion of the galaxy relative to the dense ICM removes the cold interstellar gas in the disc that is the fuel for the star formation (Gunn \& Gott 1972). Vollmer et al. (2006) modelled the heavily ram-pressure-stripped galaxy NGC 4522 in the Virgo Cluster (Kenney, van Gorkom \& Vollmer 2004). The ram-pressure-stripping model successfully reproduces the observed H I gas deficiency and the truncated gas disc of the galaxy (Vollmer et al. 2006). The time-scale for the ram-pressure stripping to terminate star formation was estimated to be 0.01-0.1 Gyr from numerical simulations (Abadi et al. 1999; Quilis et al. 2000; Fujita \& Nagashima 1999).

Some authors, however, argue that ram pressure alone cannot explain the observed decline in star formation in cluster galaxies (Treu et al. 2003; Balogh et al. 2002; Kodama \& Bower 2001). Treu et al. (2003) reported the mild decline in star formation at the periphery of the cluster Cl 0024$+$16, where the ram pressure may not be effective in stripping cold disc gas.

Other possible mechanisms have been proposed that require relatively longer time-scales to strip the galactic halo gas (Larson, Tinsley \& Caldwell 1980), and these have been termed 'strangulation' (Fujita 2004; Tanaka et al. 2004) or 'starvation' (Treu et al. 2003; Boselli et al. 2006; Bekki et al. 2002). 'Strangulation' involves the stripping of warm halo gas through the interaction with the ICM in a situation where further supply of halo gases to the disc is disrupted. Because this can occur even in less dense environments, this may explain the observed decline of star formation among galaxies in the outer regions of clusters.

Mechanisms such as mergers cannot be responsible for creating passive spiral galaxies because they may disturb the spiral arms. Similarly, 'harassment', high-speed gravitational encounters between galaxies (Moore et al. 1996), may also lead to changes in morphology and thus cannot be a dominant mechanism for transforming normal spiral galaxies into passive ones.

In this paper, we perform spatially resolved spectroscopy on four of the passive spiral galaxies identified in our previous paper (Goto et al. 2003b) in order to obtain further constraints on the mechanism responsible for halting the star formation. The strength of the H$\delta$ absorption line and 4000-{\AA} break are compared with the simple stellar population (SSP) model constructed by Bruzual \& Charlot (2003) in order to estimate the age of the stellar population. We then attempt to estimate the time-scale for the mechanism to create passive spiral galaxies.

The method of data reduction and analysis are described in Section 2, the results are in Section 3, the discussion based on the comparison with the SSP model is presented in Section 4, and conclusions are given in Section 5. Unless otherwise stated, we adopt the best-fitting WMAP cosmology: (h, $\Omega_{m}$, $\Omega_{L}$) = (0.71, 0.27, 0.73) (Bennett et al. 2003).

\begin{table*}
 \centering
 \begin{minipage}{140mm}
 \label{tab1:1}
  \begin{tabular}{@{}lllcccccc@{}}
  \hline
    & & & & & & \multicolumn{2}{c}{Signal-to-noise ratio\footnote{Signal-to-noise ratio measured by taking the standard deviation of the continuum waveband defined for spectral indices}}\\
      Name & Ra(J2000)&Dec(J2000)&$z$&$\sigma_{V}$ km s$^{-1}$\footnote{Velocity dispersion $\sigma_{V}$ and Petrosian radius $r_{\rm P}$ in $r'$ band are taken from SDSS Data Release 5 \citep{b3,Adelman}  }&$r_{\rm P}$ arcsec & ${\rm AP}_{in}$&${\rm AP}_{out}$ & $M_r$\footnote{Absolute magnitude in $r'$ band}\\
 \hline 
SDSS J021534.35$-$090537.0 & 02 15 34.35&$-$09 05 37.0&0.0687 & 112$\pm$7&15.62 &19.7& 9.8& $-$22.58\\
SDSS J024732.02$-$065137.5 & 02 47 32.02&$-$06 51 37.5&0.0705 &131$\pm$9 &10.31 & 17.2&12.0& $-$21.75\\
  SDSS J033322.66$-$000907.5&03 33 22.66&$-$00 09 07.5&  0.0838 & - & 6.615 & 11.1 & 7.0 & $-$20.89\\ 
 SDSS J074452.51$+$373852.7&07 44 52.51&$+$37 38 52.7&  0.0743 &94$\pm$9& 8.329 & 9.4 & 4.6 &  $-$20.80\\
\hline
\end{tabular}
 \caption{Properties of target galaxies. Aperture radii for the spatially resolved analysis ($r_{\rm P}$) are also shown.}
\end{minipage}
\end{table*}

\begin{figure*}
\begin{minipage}{140mm}
\includegraphics[width=60mm,clip]{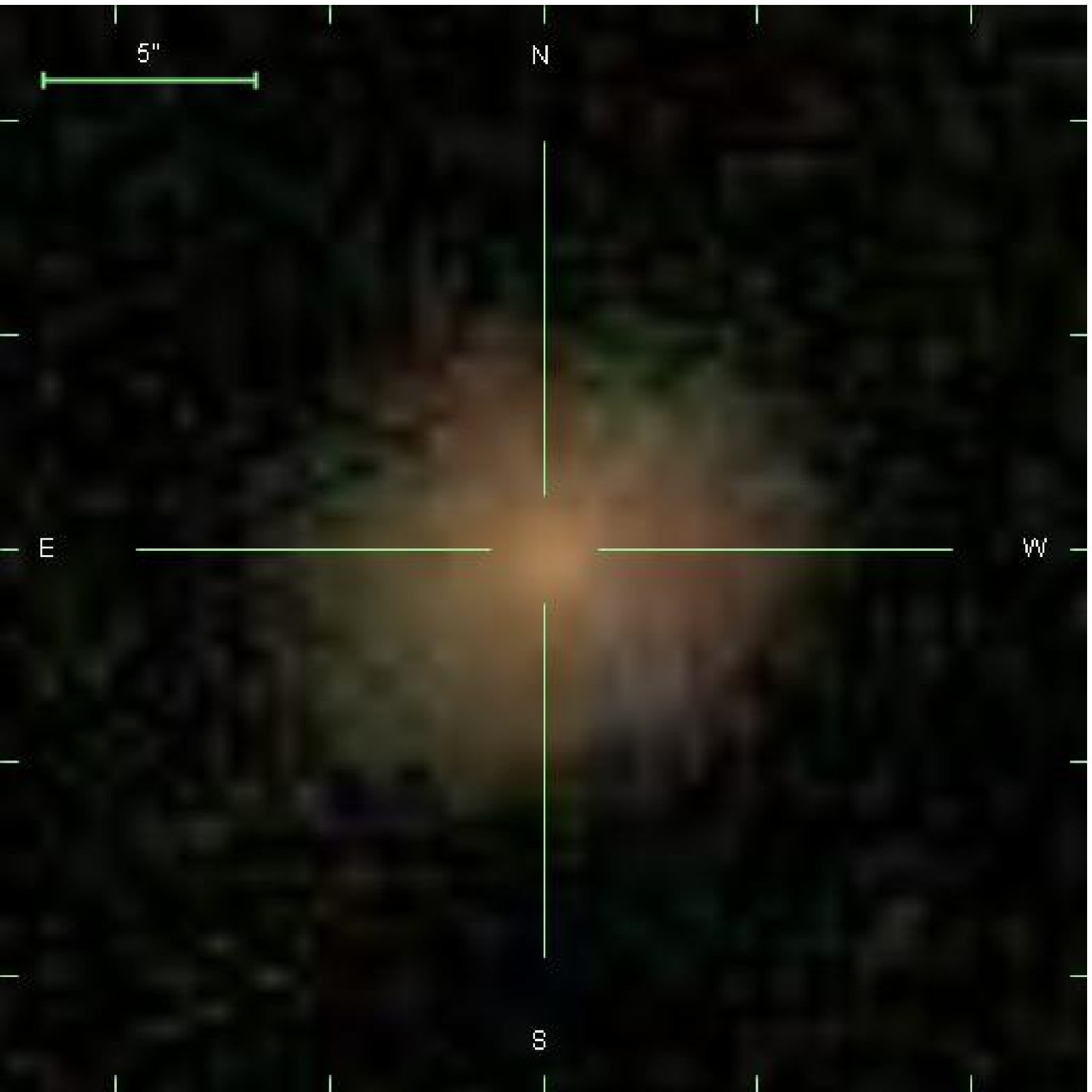}
\includegraphics[width=60mm,clip]{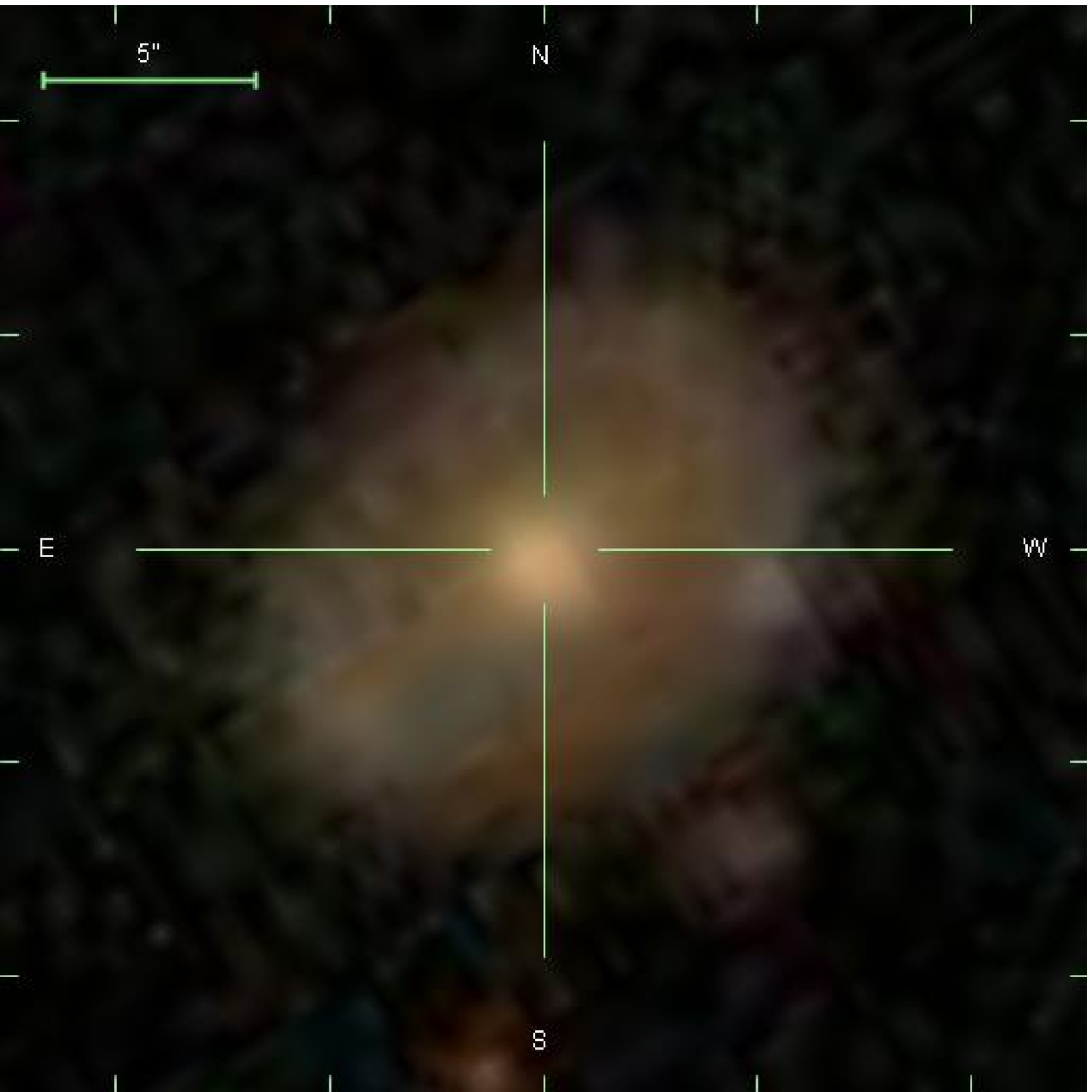}
\caption{SDSS $g', r', i'$ composite images \citep{Fukugita} of target galaxies (left: SDSS J033322.65$-$000907.5, right: SDSS J074452.52$+$373852.7)}
\label{SDSSpictures}
\end{minipage}
\end{figure*}

\section[]{The Method}
\label{sec:section2}
\subsection{Sample selection}
The target galaxies, SDSS J021534.35$-$090537, J024732.02$-$065137.5, J033322.65$-$000907.5 (the left panel of Fig. 1) and J074452.52$+$373852.7 (the right panel of Fig. 1), are a subset of the passive spiral galaxies selected in Goto et al. (2003b) from the volume-limited sample of the SDSS data. Galaxies are selected based on the following criteria: (1) the inverse concentration parameter, which is defined as the ratio of the Petrosian 50 per cent radius to the Petrosian 90 per cent radius, is less than 0.5 (details of the use of this parameter for classifying morphological type are given in Shimasaku et al. 2001) in order to select galaxies having spiral morphology; (2) the absence of [O II] and H$\alpha$ emission lines (the measured values of equivalent width are less than 1 $\sigma$ error, Goto et al. 2003b), which are indicative of ongoing star formation activity.

\subsection{Observation and Data reduction}
\label{subsec:2}
The observations were carried out using the Dual Imaging Spectrograph (DIS) installed on the Apache Point Observatory 3.5-m telescope on 2004 October 18. Both the blue and red cameras on the DIS were used in a medium-dispersion mode, with the dispersion covering the wavelength range 3000-9000 {\AA}. The pixel scales of the spatial axis for the blue and the red cameras were 0.42 and 0.40 arcsec pix$^{-1}$, respectively. To perform spatially resolved spectroscopy, a long slit with a slit width of 1.5 arcsec was used. Each sample galaxy was observed three times with an exposure time of 1000-1500 s. After bias-subtraction and flat-fielding had been applied, the three frames were combined into one frame. Spectra of standard stars, HR 718 and Hilt 600, were taken with an exposure time of 20 and 1 s, respectively, and used for flux calibration. The seeing size measured using the full width at half maximum (FWHM) of the point spread function of Hilt 600, which was monitored during the observation of the target galaxies, was $\sim$1.7 and $\sim$1.3 arcsec for the red and the blue camera, respectively.

To perform the spatially resolved analysis, we refer to the Petrosian radius in r' band ($r_{\rm P}$), which is a measure of the surface brightness profile of galaxies, obtained through the SDSS (Blanton et al. 2001). The values are presented in Table 1. The long-slit data were divided into 11 spatial bins using the iraf routine apall, which outputs 11 spectra for all bins. Then, the three bins around the centre and the remaining eight bins sampling the spectra of two sides of the galaxies were summed, respectively, to increase the signal-to-noise ratio. Hereafter, these summed data for the inner and outer parts of the galaxies are designated as AP$_{in}$ and AP$_{out}$, respectively. AP$_{in}$ samples approximately $r$ $\leq$ 0.27$r_{\rm P}$, whereas AP$_{out}$ samples 0.27$r_{\rm P}$ $<$ r $\leq$ 1$r_{\rm P}$, where $r$ represents the distance from the galaxy centre. The diameter of the inner part, $\sim$3.7 arcsec, is much larger than the seeing size ($<$1.7 arcsec). Wavelength calibration was performed using observations of a HeNeAr lamp. The sensitivity function was obtained using the spectra of the standard stars HR 718 and Hilt 600. Because it was difficult to fit the whole wavelength range with a single sensitivity function, we excluded the data points at the edges of each red and blue spectrum. The resulting spectral coverage, included in the fitting of the sensitivity correction, is 3700-5650 {\AA} for the blue spectrum and 6300-9500 {\AA} for the red spectrum.

The wavelength resolution was measured using the FWHM of the emission lines of the HeNeAr lamp spectra. The result was $\sim 8.5 \pm 0.5$ {\AA} FWHM at approximately 5000 {\AA}.

\subsection{Measurement of spectral features}
\label{subsec:3}
\begin{figure*}
\begin{center}
\label{fig:ha}
\includegraphics[width=45mm,angle=270]{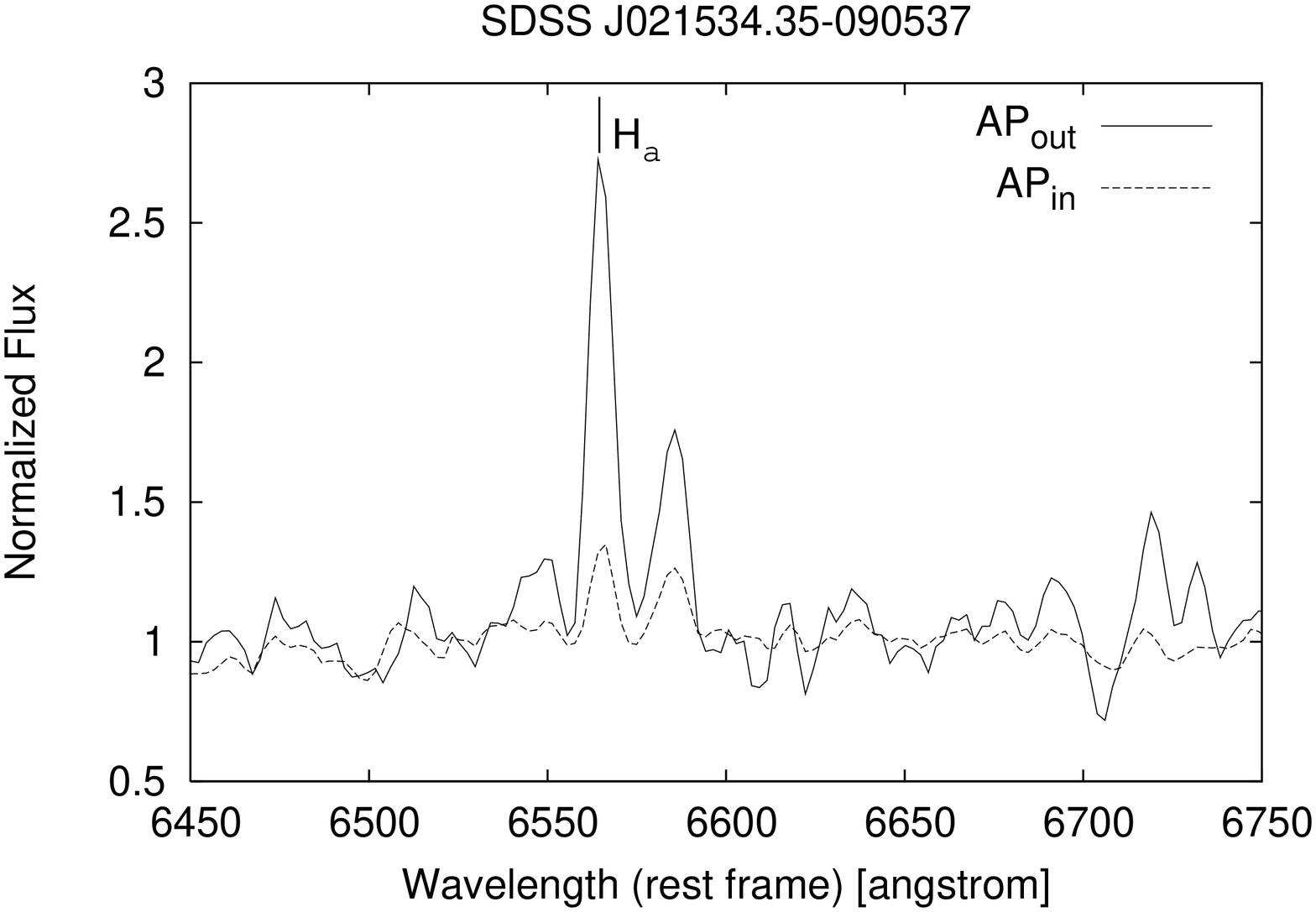}
\includegraphics[width=45mm,angle=270]{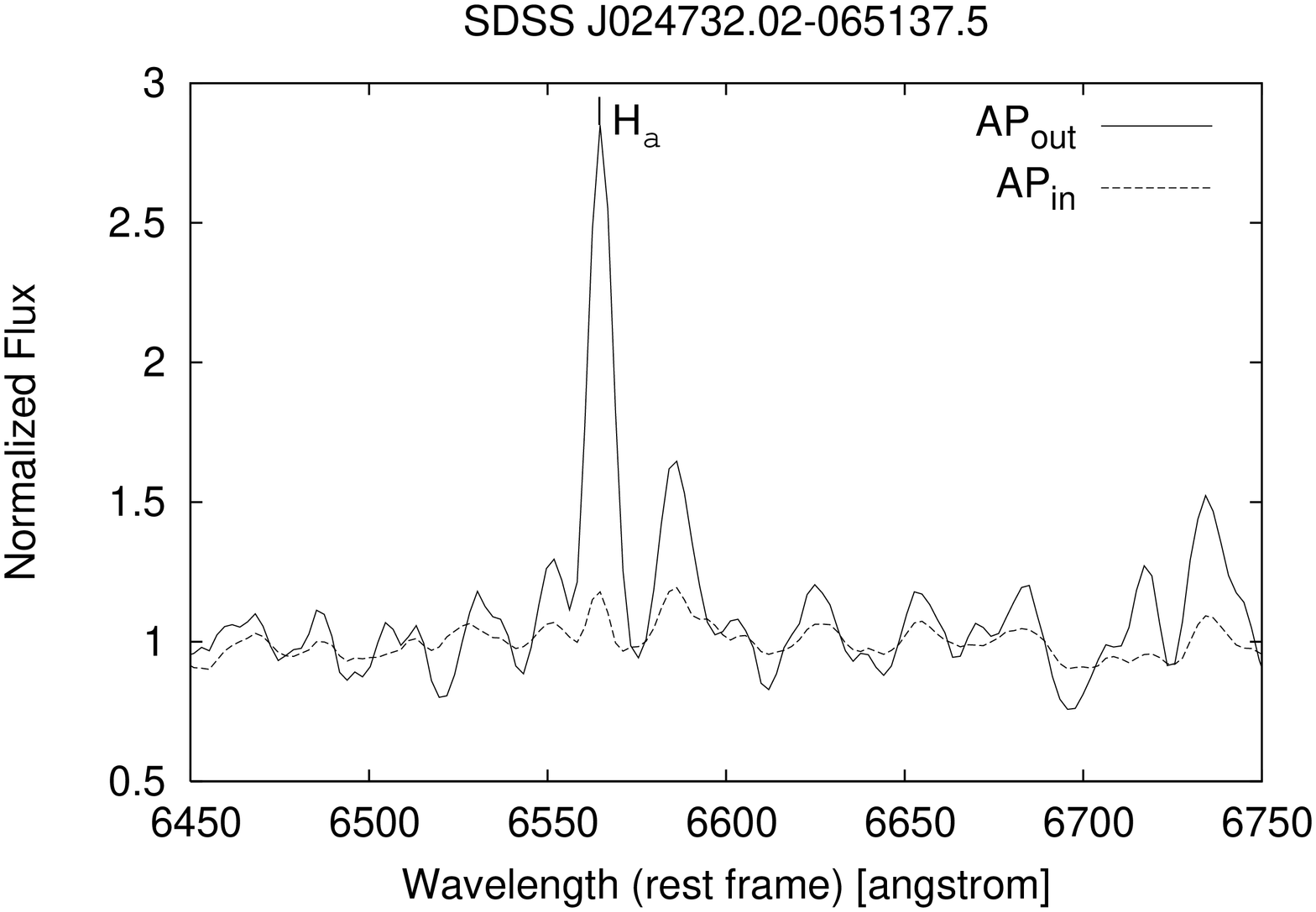}
\includegraphics[width=45mm,angle=270]{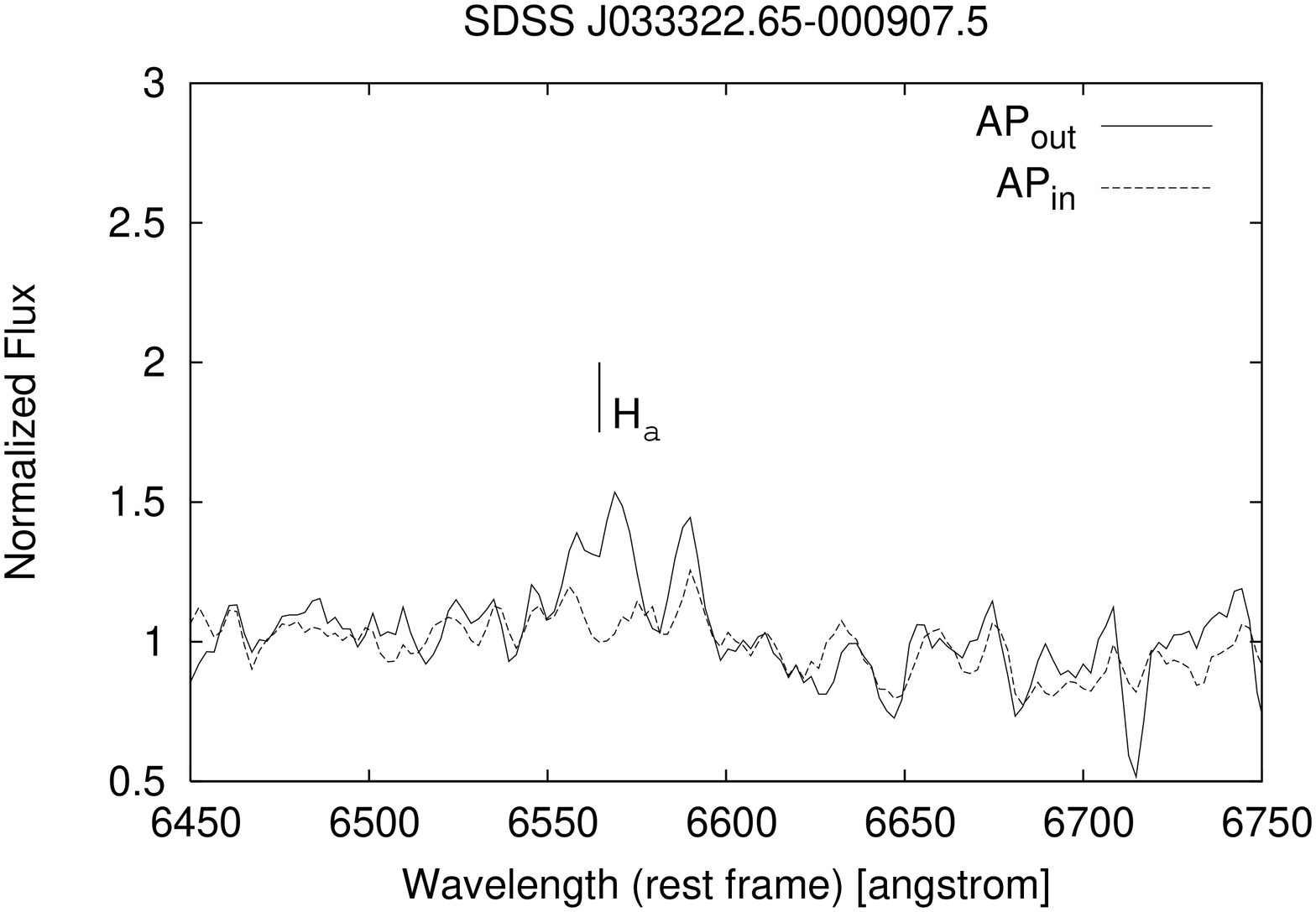}
\includegraphics[width=45mm,angle=270]{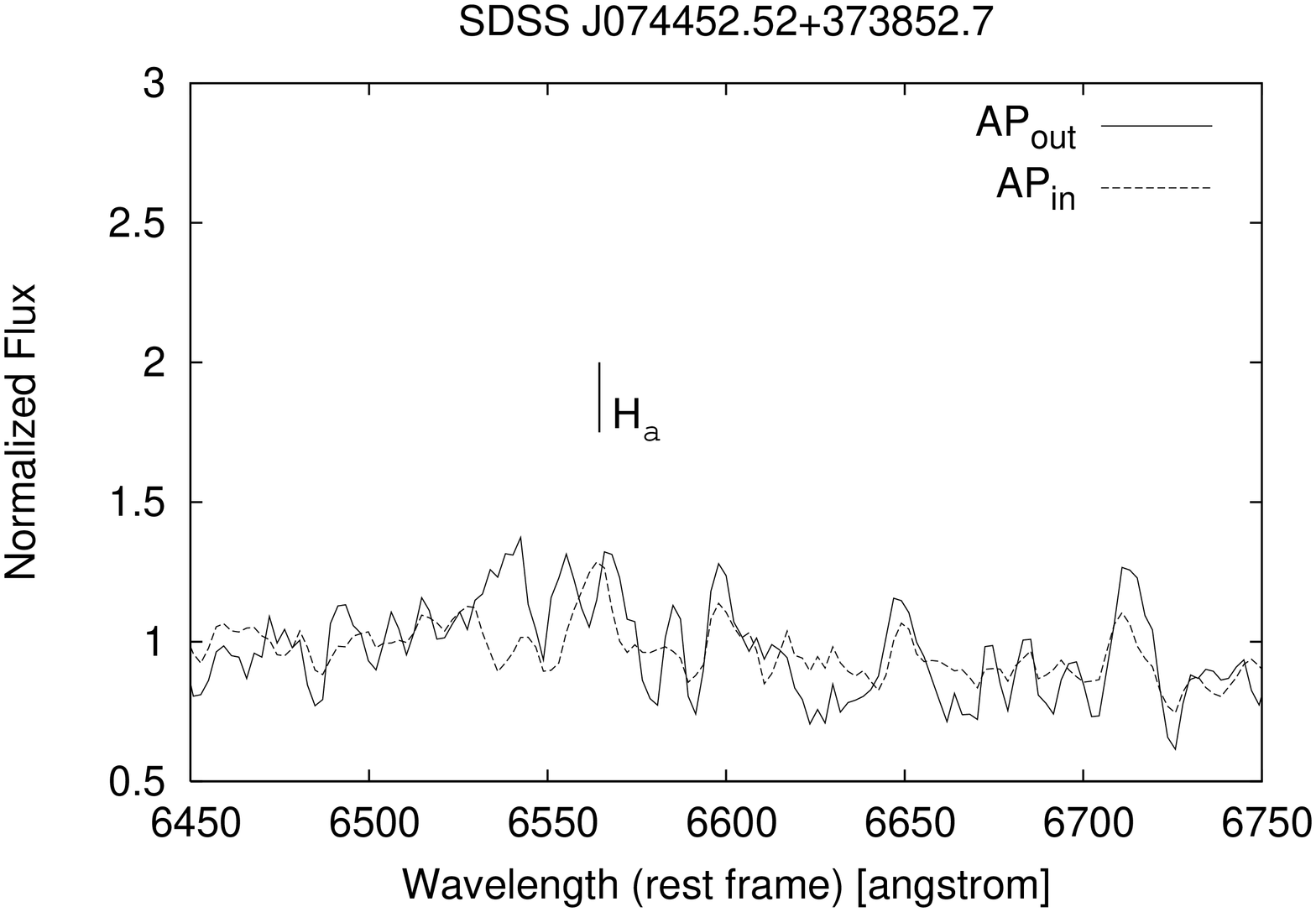}
\caption{The rest-frame spectra around the ${\rm H}\alpha$ emission line for the observed galaxies. Solid lines show the spectra for ${\rm AP}_{out}$ and dotted lines show the spectra for ${\rm AP}_{in}$.  }
\end{center}
\end{figure*}

Before we measured the spectral features, the redshift for each galaxy was computed from the Ca H line at 3970 {\AA}, one of the most prominent features in the obtained spectra, by fitting a Gaussian for this line with the iraf splot routine. The wavelengths of the observed spectrum were shifted to the rest-frame based on the derived redshifts. The derived redshifts are shown in Table 1.

The Lick/IDS absorption-line index system (Trager et al. 1998; Worthey 1994) was used to measure the strengths of absorption features in the spectra. The wavelength definitions for the index measurement were taken from Worthey \& Ottaviani (1997) for the H$\gamma$ and H$\delta$ indices, and from Trager et al. (1998) for the other 21 indices. The equivalent width or magnitudes for each index were computed as defined in Trager et al. (1998). We also measured the 4000-{\AA} break (D$_{4000}$) to estimate the age of the stellar population, as it is a broad feature and can be measured with a greater signal-to-noise ratio than can individual lines. D$_{4000}$ is widely used as a diagnostic of the age and metallicity of stellar populations and can be compared with values in the literature (Bruzual 1983; Gorgas et al. 1999; Kauffmann et al. 2003). D$_{4000}$ is calculated as defined in Balogh et al. (1999), who take a narrower spectral region for the red and blue continua than do Bruzual (1983). The reason for using this narrower definition is that it is less affected by the uncertainty in the sensitivity correction, and, more importantly, by the reddening effect (Balogh et al. 1999).

In order to confirm the absence of star formation activity for the observed candidates of passive spiral galaxies, equivalent widths of H$\alpha$ and [O II] emission line were measured. Fig. 2 shows the spectral regions around the H$\alpha$ emission lines. Two of the four target galaxies (SDSS J021534.35$-$090537 and SDSS J024732.02$-$065137.5) are found to show detectable H$\alpha$ emission (H$\alpha$ $-$ 1$\sigma$ error $\geq$ 10 {\AA}) at AP$_{out}$, and thus do not meet the criteria for passive spiral galaxies defined in Goto et al. (2003b). The [O II] emission lines, whose equivalent widths are less than 7 {\AA}, were also detected at AP out, indicating the presence of current star formation activity in the exterior regions of these galaxies. The non-detection of [O II] in the SDSS data was presumably the result of the aperture of the SDSS spectroscopic fiber (diameter of 3 arcsec), which samples only the inner parts of the galaxies, where the observed light may be dominated by the bulge component over the disc component (Abazajian et al. 2005). The results indicate the importance of using the whole light, including that from the outer regions of the galaxies, when identifying and investigating passive spiral galaxies.

We restrict our discussion to galaxies with no prominent star formation activity over the whole galaxy. The remaining two galaxies, SDSS J033322.65$-$000907.5 and SDSS J074452.52$+$373852.7, are hereafter denoted as SDSSJ0333$-$0009 and SDSSJ0744$+$3738. The velocity dispersion (available only for SDSSJ0744$+$3738), Petrosian radius, absolute magnitude in the r' band obtained by the SDSS, and the measured values of redshift and signal-to-noise ratio are shown in Table 1. The g', r', i'-composite SDSS images (Fukugita et al. 1996) of SDSSJ0333$-$0009 and SDSSJ0744$+$3738 are shown in Fig. 1.

\section{Results: Spectral characteristic}
\label{sec:section3}
Figs 3 and 4 show the spectra for AP$_{in}$ (upper figure) and AP$_{out}$ (lower figure) for each target galaxy over the rest-frame wavelength range of 3900-4900 {\AA}. These spectra are smoothed by a boxcar function over two data points and normalized to the flux at 4400 {\AA} to clarify the comparison of the absorption strengths. In the wavelength bands of the major spectral indices, H$\delta$, Ca, G4300, H$\gamma$, Fe and H$\beta$ are marked. Smoothed spectra around the D$_{4000}$ and the H$\delta$ absorption are shown in Fig. 5. To enable visualization of the strength of D$_{4000}$, average fluxes over the red (4000-4100 {\AA}) and blue (3850-3950 {\AA}) bandpasses are represented by horizontal lines. Table 2 summarizes the measured equivalent widths of H$\delta_{\rm A}$, H$\delta_{\rm F}$ (the subscripts A and F refer to a wider and narrower index definition, respectively) and D$_{4000}$. It also shows these indices measured for the spectrum obtained in SDSS DR5 (Adelman-McCarthy et al. 2006) using the same bandpasses and methodology as are applied to our samples. The errors in the indices are computed by propagating the standard errors of the pseudo-continuum bandpasses. Errors for D$_{4000}$ are estimated in such a way that the errors of each pixel propagate according to the definition of D$_{4000}$ (Balogh et al. 1999).

The following subsections describe the characteristics of the spectra for each passive spiral galaxy.

\subsection{SDSS J033322.65$-$000907.5}
\begin{figure}
\begin{center}
\caption{The rest-frame spectra for AP$_{out}$ and AP$_{in}$ for SDSS J033322.65$-$000907.5. The vertical axis shows the flux normalized to the flux at 4400 {\AA}. The horizontal axis shows the wavelength in the rest-frame. The spectra were smoothed with a boxcar function of width two data points. Shaded areas show wavebands used for the measurement of the spectral indices as defined in Worthey \& Ottaviani (1997) and Trager et al. (1998).}
\label{fig:spec1}
\includegraphics[width=75mm]{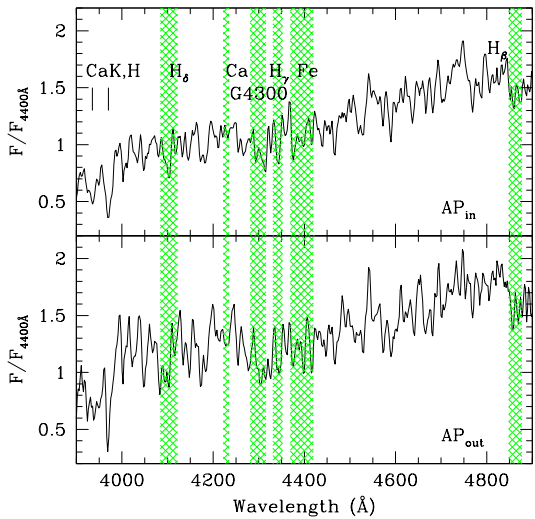}
\end{center}
\end{figure}

As shown in Fig. 3, Balmer absorption lines, Ca and Fe absorption lines and deep absorption in G4300 bands are observed for both AP$_{out}$ and AP$_{in}$ in SDSSJ0333$-$0009. The [O II] emission line was not detected for either AP$_{out}$ or AP$_{in}$, which ensures the passive nature of this galaxy. The H$\alpha$ emission line, which is also an indicator of current star formation was, however, weakly detected for AP$_{out}$ with an equivalent width of $\sim -3$ {\AA}. H$\delta_{\rm A}$ absorption is stronger for AP$_{out}$ than that for AP$_{in}$ by $\sim$1.6 {\AA} (Table 2). By contrast, the D$_{4000}$ break is larger for AP$_{in}$ than it is for AP$_{out}$ by $\sim$ 0.07 (see Fig. 5). These results suggest that there is a higher proportion of old stars in the inner than there is in the outer regions for these galaxies.

\subsection{SDSS J074452.52$+$373852.7}
As shown in Fig. 4, the spectrum of SDSSJ0744$+$3738 is dominated by the Balmar absorptions. The metal absorptions and the deep absorption in the G4300 band look similar to those of SDSSJ0333$-$0009, but the difference in the equivalent widths of observed features between AP$_{in}$ and AP$_{out}$ is larger: H$\delta_{\rm F}$ absorption is stronger for AP$_{out}$ than for AP$_{in}$ by $\sim$4.0 {\AA} as shown in Table 2, and the H$\beta$ absorption is stronger for AP$_{out}$ by 3.5 $\pm$ 0.5 {\AA}. D$_{4000}$ is larger for AP$_{in}$ than for AP$_{out}$ by $\sim$0.5 (Table 2). These results indicate that the stellar population is younger in the outer regions, similar to the situation for the previous galaxy. The equivalent width of the H$\alpha$ emission line is found to be weak: $\sim -2.3 \pm 0.5$ {\AA} for AP$_{in}$ and $\sim -3.4 \pm 0.6$ {\AA} for AP$_{out}$. The weakness of the H$\alpha$ emission lines suggests the lack of prominent star formation activity in both the inner and the outer regions of this galaxy.

\begin{figure}
\begin{center}
\caption{The rest-frame spectra for AP$_{out}$ and AP$_{in}$ for SDSS J074452.52$+$373852.7. The notation is the same as in Fig. 3}
\label{fig:spec2}
\includegraphics[width=75mm]{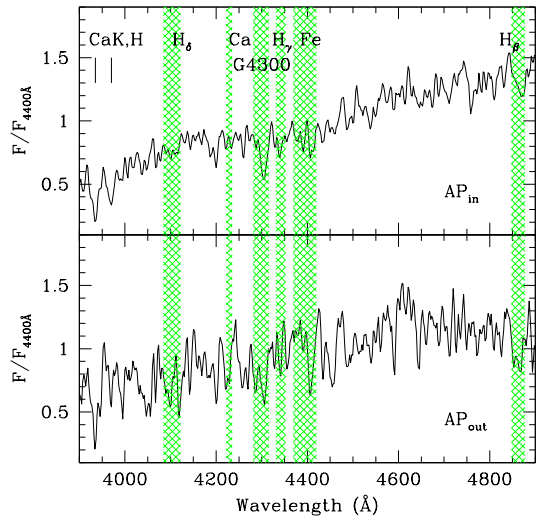}
\end{center}
\end{figure}

\begin{figure*}
\begin{center}
\includegraphics[width=55mm,angle=270]{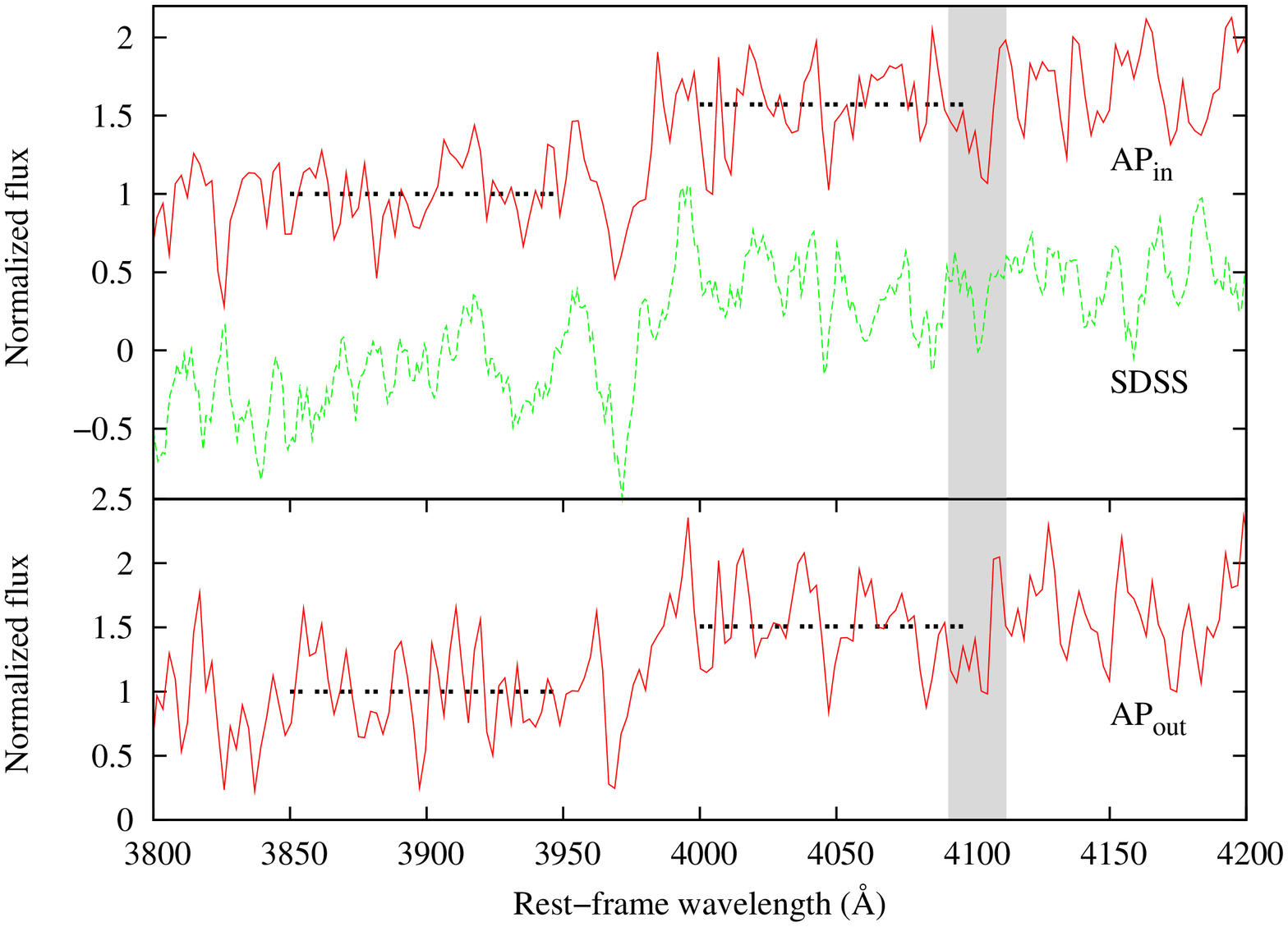}
\includegraphics[width=55mm,angle=270]{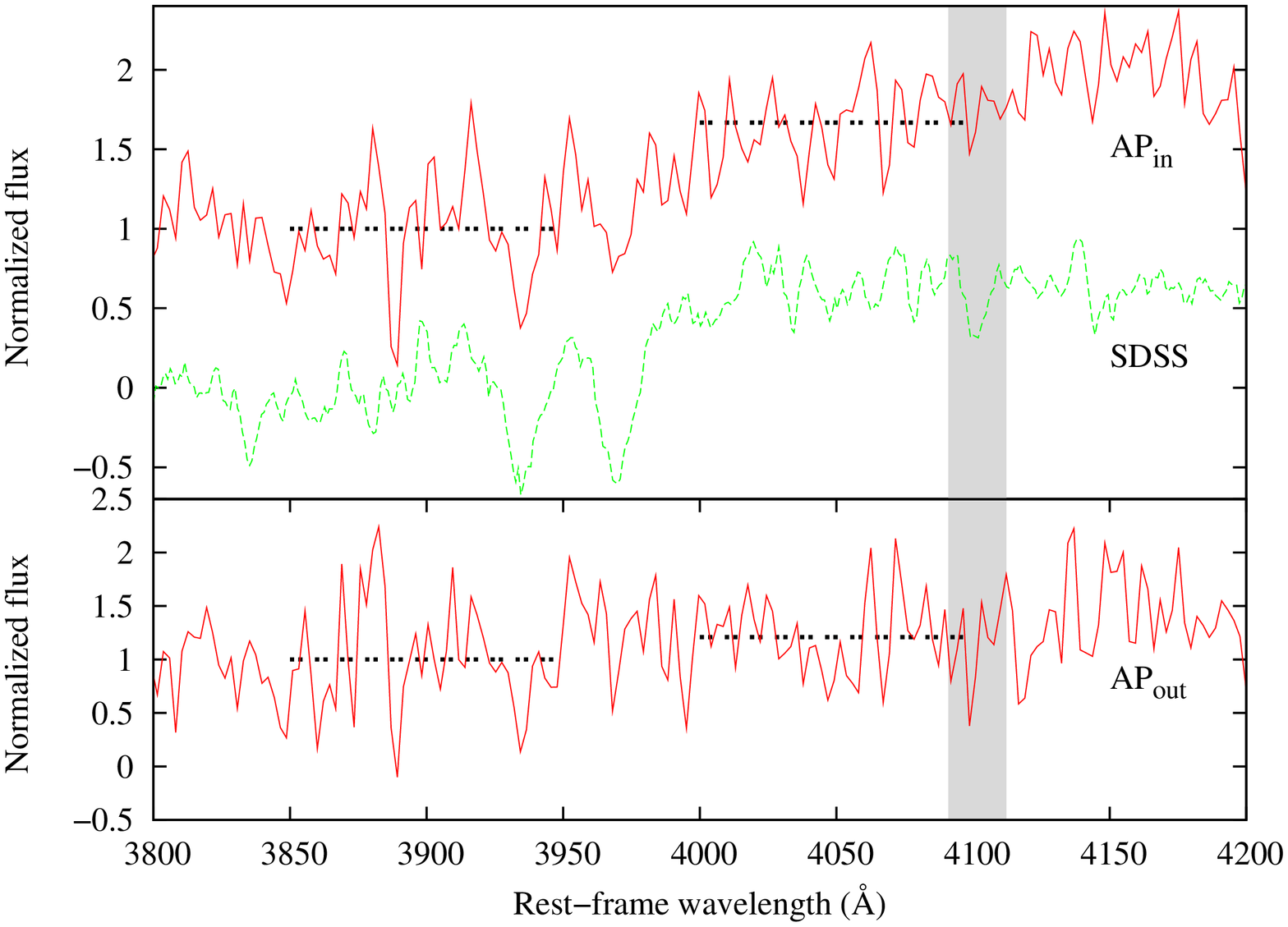}
\caption{The rest-frame spectra around the 4000-{\AA} break and the H$\delta$ absorption feature. The lower panels show the spectra for AP$_{out}$ and the upper panels show the spectra for AP$_{in}$. Below the spectrum for AP$_{in}$, spectra obtained from the SDSS are shown offset from the observed spectrum for comparison. Horizontal lines show the average fluxes over the red (4000-4100 {\AA}) and the blue (3850-3950 {\AA}) continuum bandpasses. The spectra and the average fluxes are normalized to the average fluxes over the blue continuum bandpasses. The shaded areas show the bandpasses of the H$\delta_{\rm F}$ index.}
\label{fig:d4000}
\end{center}
\end{figure*}

\begin{table*}
\caption{Equivalent widths of H$\delta_{\rm A}$, H$\delta_{\rm F}$ and the values of D$_{4000}$ are shown. The last three columns show the indices measured for the spectra obtained from the SDSS (Adelman-McCarthy et al. 2006) with the same methodology as used for our samples.}
\label{table2}
\begin{tabular}{@{}lccccccc@{}}
\hline
Target & Spatial bin &${\rm H}\delta_{A}$ &${\rm H}\delta_{F}$& $D_{4000}$&${\rm H}\delta_{A}$(SDSS)& H$\delta_{F}$(SDSS)&$D_{4000}$(SDSS)\\ \hline
SDSS J033322.65$-$000907.5 &${\rm AP}_{in}$ &2.0$\pm$1.8 &2.6$\pm$1.1&1.68$\pm$ 0.03 &-1.9$\pm$2.4 & 0.2$\pm$1.6 &1.66 \\ 
 &${\rm AP}_{out}$&3.6$\pm$4.3&4.4$\pm$ 2.6 &1.61$\pm$0.03 & & &\\ 
SDSS J074452.52$+$373852.7 &${\rm AP}_{in}$ &1.0 $\pm$1.7 &1.6$\pm$1.2& 1.80$\pm$ 0.03 &-0.1$\pm$1.2 & 1.1$\pm$0.7  &1.72  \\ 
 & ${\rm AP}_{out}$ & 22.2$\pm$8.1 &5.5$\pm$11.0&  1.29$\pm$0.02 &&&\\
\hline
\end{tabular}
\end{table*}

\section{Discussion}
\label{sec:section4}
In the following subsection, we first try to estimate the metallicity and the effects of $\alpha$-enhancement on the absorption strengths for the observed galaxies using the measured Lick indices. The super-solar $\alpha$/Fe ratio could affect the strength of the H$\delta_{\rm A}$ and the H$\delta_{\rm F}$ indices, as discussed in Thomas, Maraston \& Bender (2003), which leads to an underestimate of the age of stellar populations. Considering the estimated metallicity and the effect of $\alpha$-enhancement, Section 4.2 discusses the light-averaged ages of the stellar populations using a H$\delta_{\rm F}$-D$_{4000}$ plane.

The Lick indices measured with $>$1$\sigma$ for both the inner and outer regions were Fe4383, Fe4668, H$\beta$, Fe5270 and H$\delta_{\rm F}$ for SDSSJ0333$-$0009. For SDSSJ0744$+$3738, G4300, Fe4531 and Fe4668 were detected with $>$1$\sigma$. Unfortunately, the H$\delta_{\rm F}$ index, which is used as an age indicator, does not have a good enough signal-to-noise ratio for SDSSJ0744$+$3738. Furthermore, neither the inner nor the outer region of this galaxy has an Mgb index, which is used as an indicator of metallicity and $\alpha$-enhancement, with a signal-to-noise ratio $>$1. We therefore focus our discussion of the metallicity and $\alpha$-enhancement on SDSSJ0333$-$0009, for which the iron indices and the Mgb indices were measured with a relatively high signal-to-noise ratio.

\subsection{Metallicity and $\alpha$-enhancement}
\label{sec:metal}
The measured Lick indices (Fe4383, Fe4668, H$\beta$, Fe5270 and H$\delta_{\rm F}$) for SDSSJ0333$-$0009 indicate that the inner region is more metal-rich than the outer region. Fig. 6 shows the strength of these indices for AP$_{in}$ (open triangles) and AP$_{out}$ (filled triangles) as a function of the Mgb index. The SSP models of Thomas et al. (2003) with age = 1.0, 5.0, 10.0, 15.0 Gyr and metallicity Z = 0.67, 0.00(Z$_{\odot}$), -2.25, assuming [$\alpha$/Fe] = 0.0, are overlaid. A comparison of the strength of the observed indices with that of the SSP models suggests that the observed indices for both the inner and outer regions are consistent with the model with super-solar metallicity. The iron indices (Fe4383, Fe4668 and Fe5270), especially, suggest that there is a radial gradient in the metallicity; the observed indices for the inner region are marginally consistent with [Z/H] = 0.67 models, whereas for the outer region the observed indices are reproduced by the model with [Z/H] = 0.00 in the Fe4668-Mgb plane and [Z/H] = 0.00-0.67 in the Fe5270-Mgb plane. Therefore, although the index strengths are associated with large uncertainties, they suggest that the inner region is more metal-rich than the outer region. It should be noted that H$\delta_{\rm F}$ for both regions and Fe4383 for the outer region were not reproduced by the models for [$\alpha$/Fe] = 0.0. Rather, they seem to be consistent with models with [$\alpha$/Fe] = 0.5, as discussed in the next section.

Comparisons of the Fe4383 indices, which are especially sensitive to the $\alpha$-enhancement (Thomas et al. 2003), with the model suggest that the outer region of this galaxy possibly has a super-solar $\alpha$/Fe ratio. Fig. 7 shows the strength of the Fe4383 index as a function of the Mgb index. Models with [$\alpha$/Fe] = 0.0 and 0.5 for ages and metallicities the same as those shown in Fig. 6 are overlaid. Open and filled triangles show the observed index strength for AP$_{in}$ and AP$_{out}$, respectively. This figure indicates that the strength of Fe4383 for AP$_{out}$ is more consistent with the models with [$\alpha$/Fe] = 0.5 than it is with those with [$\alpha$/Fe] = 0.0.

In order to elucidate the dependence of the H$\delta$ absorption strength on [$\alpha$/Fe], Fig. 8 shows the model predictions for the H$\delta_{\rm F}$ index as a function of the Mgb index for [$\alpha$/Fe] = 0.0 and 0.5. This illustrates that the difference in H$\delta_{\rm F}$ absorption owing to the $\alpha$-enhancement is expected to be $\sim$ 0.8-1.2 {\AA} , depending on the age and metallicity of the stellar population.

The inner region (open triangles) is roughly consistent with the model with age $\sim$ 1.0-1.5 Gyr, metallicity Z $\sim$ 0.67, and [$\alpha$/Fe] = 0.5. The outer region (filled triangles), by contrast, shows stronger H$\delta_{\rm F}$ absorption, which suggests that the stellar population in the outer region is, on average, younger than that in the inner region. The age and the metallicity implied from the plot for H$\delta_{\rm F}$ against Mgb are $<$ 1.0 Gyr and Z $\sim$ 0.00-0.67, respectively. Although the metallicity may be difficult to determine quantitatively from Fig. 8 because of the quite large uncertainty in the Mgb index, Z $\sim$ 0.00-0.67 is consistent with the value implied from the strength of the Fe4668 and Fe5270 indices against the Mgb index as shown in Fig. 6.

\begin{figure*}
  \begin{center}
    \begin{tabular}{ccc}
      \rotatebox{270}{\resizebox{38mm}{!}{\includegraphics{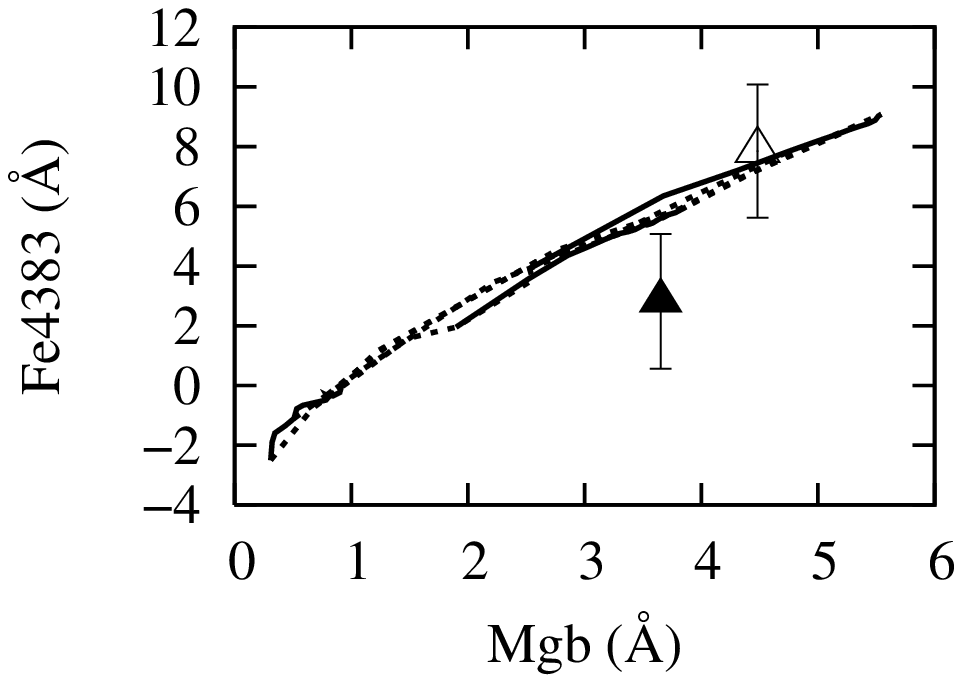}}} &
      \rotatebox{270}{\resizebox{38mm}{!}{\includegraphics{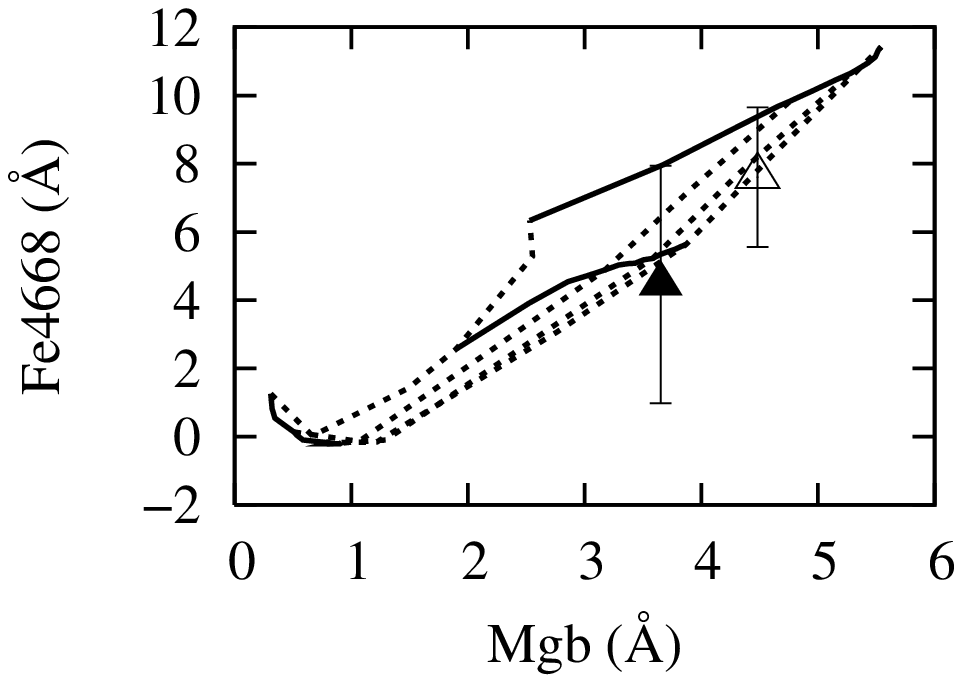}}} &
 \rotatebox{270}{\resizebox{38mm}{!}{\includegraphics{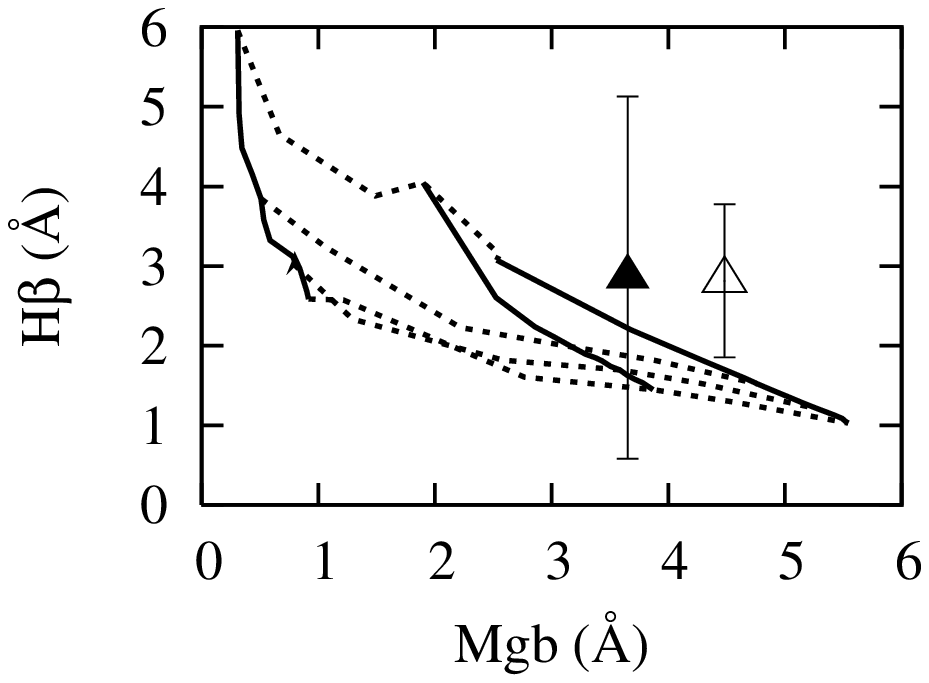}}} \\
 \rotatebox{270}{\resizebox{38mm}{!}{\includegraphics{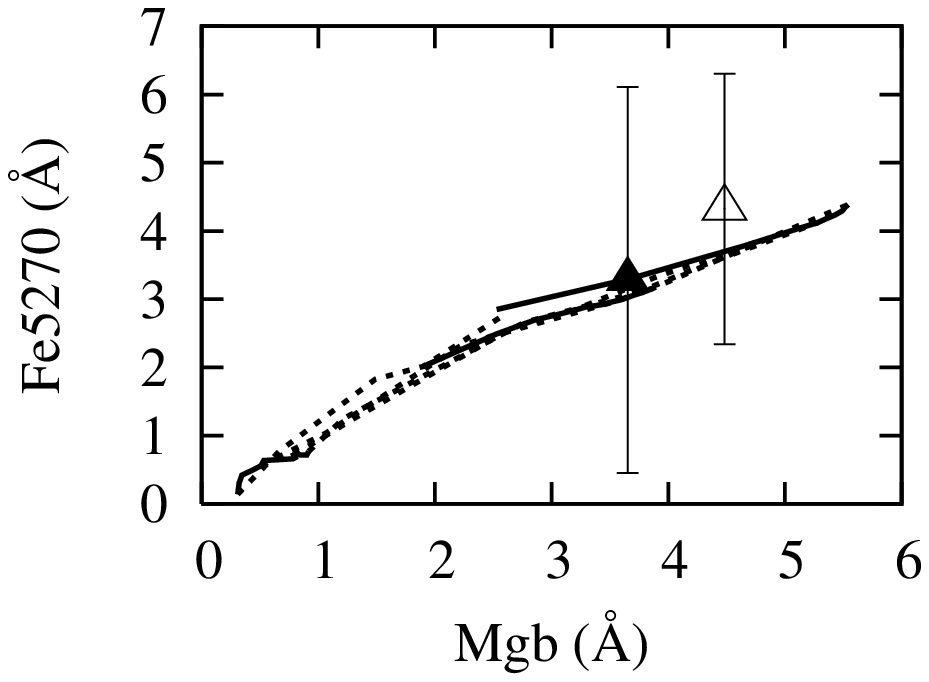}}} &
 \rotatebox{270}{\resizebox{38mm}{!}{\includegraphics{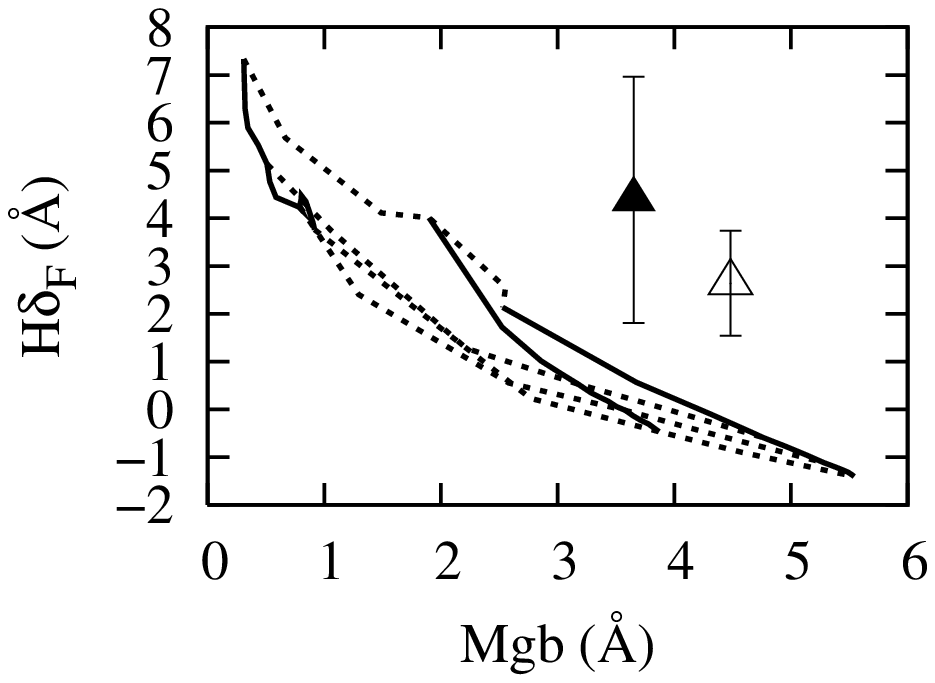}}} &   \\ 
\end{tabular}
    \caption{he Lick indices measured with $>$1$\sigma$ as a function of the Mgb index for SDSSJ0333$-$0009. Open and filled triangles show the measured index strength for AP$_{in}$ and AP$_{out}$, respectively. The solid lines show the models with [Z/H] = $-2.25$, $0.00$(Z$_{\odot}$) and 0.67 (from left to right). The dotted lines show the models with ages = 1, 5, 10 and 15 Gyr (for the Fe4383, Fe4668 and Fe5270 indices, the age increases from left to right; for the H$\beta$ and H$\delta_{\rm F}$ indices, the age increase, from top to bottom).}
    \label{mgb_metals}
  \end{center}
\end{figure*}

\begin{figure*}
\begin{tabular}{c@{\hspace{1cm}}c}
\begin{minipage}{0.45\hsize}
\begin{center}
\includegraphics[width=55mm,angle=270]{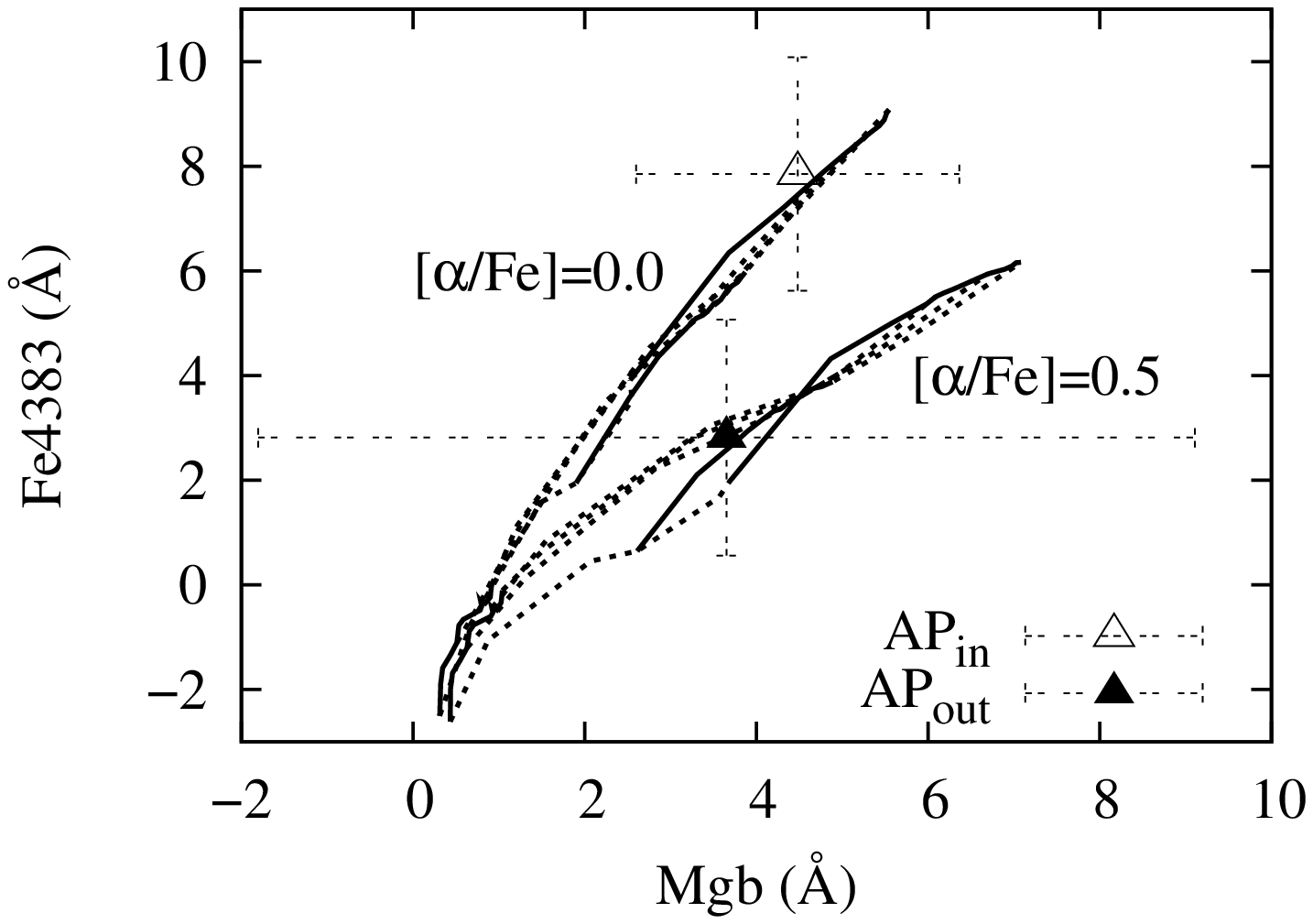} 
\end{center}
\caption{The Fe4383 index for AP$_{in}$ and AP$_{out}$ as a function of the Mgb index. The models with [$\alpha$/H] = 0.0 and 0.5 are overlaid. For both values of [$\alpha$/H], the solid lines show the models with [Z/H] = $-2.25$, 0.00 and 0.67 (from bottom to top) and the dotted lines show the models with ages = 1, 5, 10 and 15 Gyr (from bottom to top)}
\label{mgb_fe4383}
\end{minipage} &
\begin{minipage}{0.45\hsize}
\begin{center}
\includegraphics[width=55mm,angle=270]{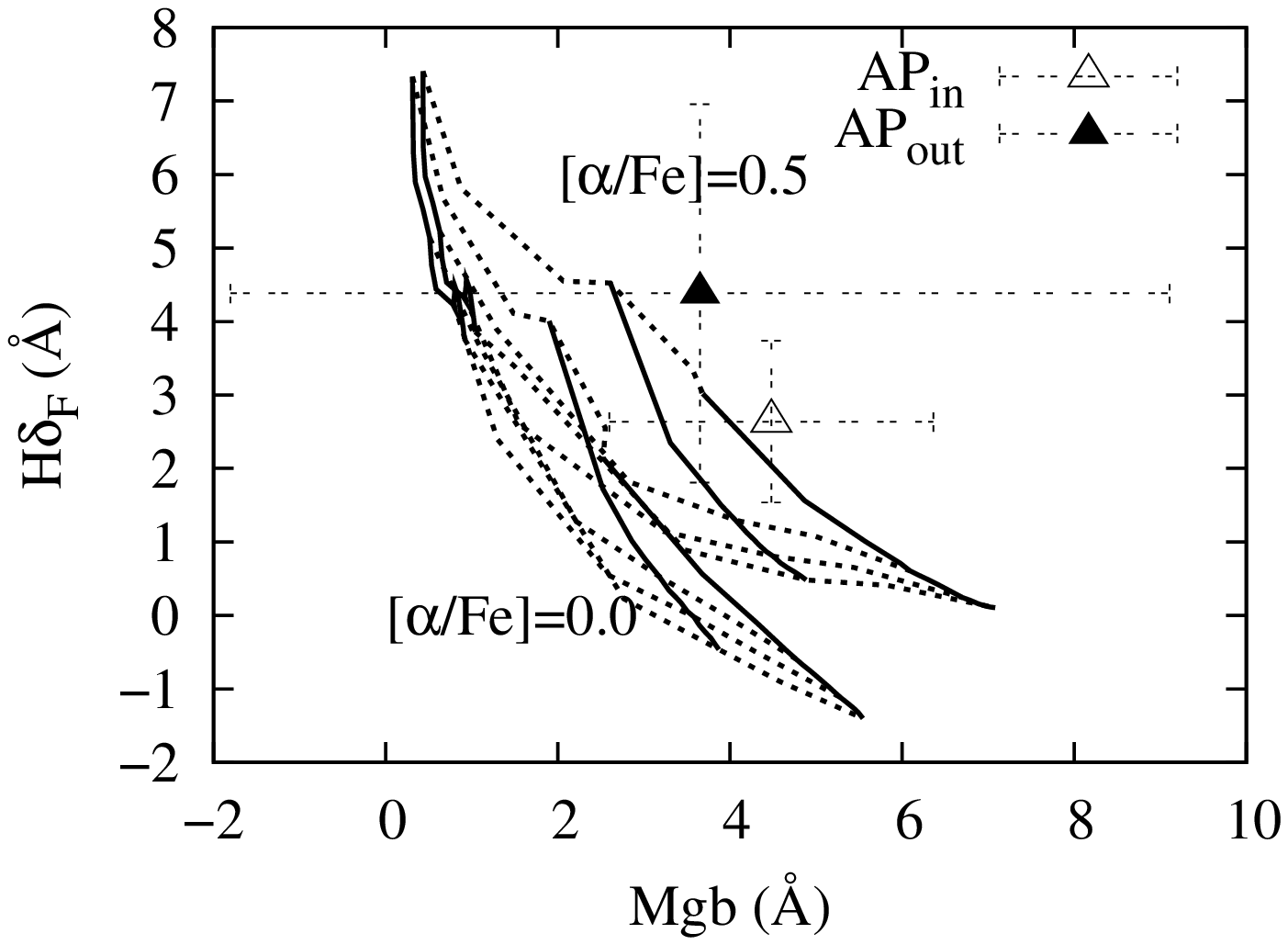}
\end{center}
\caption{The H$\delta_{\rm F}$ index for AP$_{in}$ and AP$_{out}$ as a function of the Mgb index. The models with [$\alpha$/H] = 0.0 and 0.5 are overlaid. For both values of [$\alpha$/H], the solid lines show the models with [Z/H] = $-2.25$, 0.00 and 0.67 (from left to right) and the dotted lines show the models with ages = 1.0, 5.0, 10.0 and 15.0 Gyr (from bottom to top).}
\label{mgb_hdF}
\end{minipage}
\end{tabular}
\end{figure*}

\subsection{Comparison with the SSP models}
\label{sec:age}
In the following discussions, we use the H$\delta_{\rm F}$ index and the D$_{4000}$ feature as age indicators of the stellar population. The H$\delta_{\rm F}$-D$_{4000}$ plot is widely used as a diagnostic tool to estimate the age of a stellar population in galaxies (Kauffmann et al. 2003; Poggianti \& Barbaro 1997; Balogh et al. 1999). The great advantage in using the H$\delta_{\rm F}$ and D$_{4000}$ indices as age indicators is that they are less affected by the reddening caused by interstellar dust (Kauffmann et al. 2003). We note that the ages discussed in this section refer to the light-averaged ages of stellar populations, and thus we aimed to estimate the time at which star formation ceased in the samples qualitatively.

In Fig. 9, the measured H$\delta_{\rm F}$ index and the values of D$_{4000}$ are compared with predictions from the simple stellar population (SSP) models in the galaxev package of Bruzual \& Charlot (2003). We used the model with the Salpeter initial mass function with lower and upper cut-offs of stellar masses of 0.1 ${\rm M}_{\odot}$ and 100 ${\rm M}_{\odot}$, respectively. The model assumes an instantaneous burst of star formation at age = 0 for the various metallicities (Bruzual \& Charlot 2003).

The spectra of the SSP model used in Fig. 9 were broadened to the instrumental resolution of the observations (8.5 {\AA} FWHM), which corresponds to a velocity dispersion of $\sigma$ $\sim$ 197 km s$^{-1}$. The spectral broadening was applied using the program vel\_disp included in the galaxev package (Bruzual \& Charlot 2003). The resulting SSP model tracks for metallicities [Fe/H] = $-1.6464$, $-0.6392$, $+0.0932$, $+0.5595$ are overlaid in Fig. 9. Points with the same ages (1.0, 1.6, 2.5, 4.0, 6.2, 10.0 Gyr) are joined with dotted lines.

Based on the Fe4383 index, the outer regions of SDSSJ0333$-$0009 have [$\alpha$/Fe] $>$ 0.0, as discussed in the previous subsection. Because a stellar population which has [$\alpha$/Fe] $>$ 0.0 is reported to be enhanced in H$\delta$ absorption strength compared to that having solar $\alpha$/Fe ratios with similar age (Thomas et al. 2004), we have to take this effect into account to estimate the stellar population ages using H$\delta_{\rm F}$ indices. According to the stellar population model of Thomas et al. (2003), the enhanced strength in H$\delta_{F}$ indices is up to $\sim$ 1.2 {\AA} for the model with [$\alpha$/Fe] = 0.5. Therefore, the data points should be shifted downwards from those actually plotted in Fig. 6 by up to $\sim$ 1.2 {\AA} following the correction for the $\alpha$-enhancement. For SDSSJ0744$-$3738, [$\alpha$/Fe] could not be determined from the observed spectra, because the Mg indices were not measured with a high enough signal-to-noise ratio. Therefore, the additional uncertainty arising from the $\alpha$-enhancement should be taken into account when considering H$\delta_{\rm F}$ versus D$_{4000}$ diagnostics.

Another point of caution in using the H$\delta_{\rm F}$-D$_{4000}$ plot is that the stellar absorption of H$\delta_{\rm F}$ could be contaminated by nebular emission (Kauffmann et al. 2003; Goto et al. 2003a). The nebular emission of H$\alpha$ is weak for SDSSJ0333$-$0009 and SDSSJ0744$+$3738 ($<-4$ {\AA}), and, furthermore, H$\beta$ was detected in absorption. The effect of emission filling on the H$\delta_{\rm F}$ absorption is expected to be negligible ($\sim$ 0.3 {\AA} assuming the Case B recombination value for the H$\alpha$/H$\delta$ nebulae emission line ratio without dust extinction). In the following subsection, we estimate the light-averaged age of the stellar populations in the inner and outer regions for the observed galaxies.

\subsubsection{SDSS J033322.65$-$000907.5}
The measured H$\delta_{\rm F}$ and D$_{4000}$ for AP$_{in}$ and AP$_{out}$ are shown in Fig. 9 by the open and filled triangles, respectively. If we assume a metallicity [Fe/H] $> + 0.0932$ and a correction for the $\alpha$-enhancement of $\sim$1.2 {\AA} as suggested in Section 4.1, the data for APout are broadly consistent with a model with age older than $\sim$ 1.0 Gyr. By contrast, for the inner region (AP$_{in}$) the observed H$\delta_{\rm F}$ index seems stronger than the value that can be reproduced by the model with super-solar metallicity ([Fe/H] = $+0.5595$). However, the observed strength of D$_{4000}$ for AP$_{in}$ suggests an age older than $\sim$ 1.6 Gyr. Overall, the H$\delta_{\rm F}$-D$_{4000}$ plane suggests that the average age of the stellar population is younger in the outer region than in the inner region for this galaxy.

\subsubsection{SDSS J074452.52$+$373852.7}  
\begin{figure}
\begin{center}
\includegraphics[width=80mm]{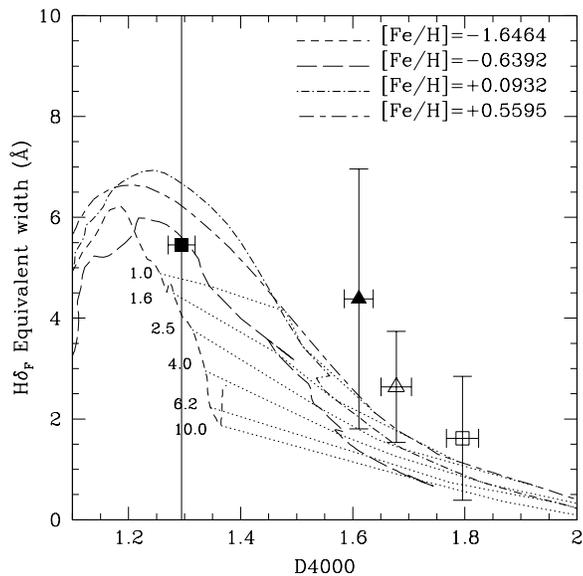}
\caption{Equivalent width of H$\delta_{\rm F}$ as a function of D$_{4000}$. The open and filled triangles represent the data for AP$_{in}$ and AP$_{out}$, respectively, for SDSSJ0333$-$0009. The open and filled squares represent the data for APin and APout, respectively, for SDSSJ0744$+$3738. Overlaid lines show predictions of SSP models broadened to the instrumental resolution ($\sigma$ = 197 km s$^{-1}$) with various metallicities [Fe/H] = $-1.6464$, $-0.6392$, $+0.0932$, $+0.5595$ (Bruzual \& Charlot 2003). Dotted lines connect points of the same age (from bottom to top, 10.0, 6.2, 4.0, 2.5, 1.6 and 1.0 Gyr, respectively).}
\label{fig:4}
\end{center}
\end{figure}

The open and filled squares in \ref{fig:4} correspond to the data for AP$_{in}$ and AP$_{out}$, respectively, for SDSSJ0744$+$3738. For this galaxy, the metallicity and [$\alpha$/Fe] are not available because of the large errors in the Lick indices. Although the determination of the age of the stellar population for SDSSJ0744$+$3738 is quite uncertain, the observed large difference in D$_{4000}$ between the core and the exterior of the galaxy imply that the two regions have distinct stellar populations in terms of their age and/or metallicity. For APin, the large value of D$_{4000}$ is marginally consistent with the model with metallicity [Fe/H] = $+0.5595$ in the H$\delta_{\rm F}$-D$_{4000}$ plane. Furthermore, in the outer region (AP$_{out}$), the H$\delta_{\rm F}$ index is quite strong (5.5$\pm$11.1{\AA}), although we note that the index strength is associated with large errors.

Galaxies with strong H$\delta$ absorption are classified as 'H$\delta$-strong' (HDS) galaxies \citep{b57,b20,b9,2004MNRAS.348..515G,b61}. As strong H$\delta$ absorption could arise from A-type stars in the main-sequence phase \citep{b2}, HDS galaxies probably terminated their starburst activity $\sim$0.1-1.5 Gyr ago \citep{b32,b9}. Therefore, the strong H$\delta_{\rm F}$ absorption in the outer regions of the observed galaxies could be clear evidence that they terminated their star formation a few Gyrs ago. Nevertheless, because of the large uncertainty associated with H$\delta_{F}$, we need further observations with higher signal-to-noise ratios to detect a clear signature of quenched star formation.

\subsubsection{Implications}
It is interesting that through the spectral diagnostics of spatially separated regions of passive spiral galaxies we can estimate the history of these galaxies. Even our spectra with relatively low signal-to-noise ratios suggest that passive spiral galaxies were in a star-forming phase for several gigayears before ceasing activity. Moreover, one of the passive spirals, SDSSJ0744$+$3738, probably terminated its star formation approximately 1-2 Gyr ago. If confirmed, this result will provide us with an important constraint in uncovering the underlying (cluster-related) physical mechanism responsible for the creation of passive spiral galaxies; for example, ram-pressure stripping \citep{b24,b37, b38,b47,b48,b25} stops star formation much quicker than the strangulation scenario \citep{b52,b46}. To reach a firm conclusion on the subject, however, further observations are needed: it is well known that the age/metallicity estimates are degenerate, for example the spectra of an old ($>$ 2 Gyr) stellar population looks almost identical when the age is doubled and the total metallicity reduced by a factor of 3 \citep{b62}. It is therefore important to break this degeneracy by obtaining spectra at a higher signal-to-noise ratio to measure indices sensitive only to age \citep[e.g. Balmer lines;][]{b4} and only to metallicity \citep[e.g.][]{b19}. Furthermore, we only had enough observing time for two passive spiral galaxies. As larger samples of passive spiral galaxies are now becoming available \citep[e.g.][]{b1}, it is important to investigate the statistical proportion of passive spiral galaxies that do not suffer from shot-noise to draw firm conclusions on this subject.

\section{Summary}
\label{sec:section5}
We have performed spatially resolved long-slit spectroscopy of four candidate passive spiral galaxies using the APO 3.5-m telescope. Two of the observed galaxies show detectable [O II] and H$\alpha$ emission lines in the exterior regions of the disc, emphasizing the importance of investigating spatially large regions including the exterior regions of passive spiral galaxies. For the other two galaxies, for which the emission lines were not detected or were weak, we found radial gradients in the H$\delta_{\rm F}$ absorption and the strength of the 4000 {\AA} break: Balmer absorption is more prominent and $D_{4000}$ is smaller in the outer regions of these galaxies. Taking into account the metallicity and the [$\alpha$/Fe] ratio roughly estimated for one of the sample galaxies, SDSSJ0333$-$0009, the comparison with the stellar population model suggests that the outer regions of the samples harbour younger populations of stars. The other observed passive spiral galaxy, SDSSJ0744$+$3738, also shows a younger population of stars in the outer regions, which presumably experienced the quenching of star formation a few Gyrs ago.

Our results have opened a door to understanding the history of cluster infall galaxies through spatially resolved spectroscopy. The results are, however, based on relatively low signal-to-noise spectra of only two passive spiral galaxies. To draw firm conclusions on this subject it is therefore essential to obtain the spectra of a larger sample of passive spiral galaxies at higher signal-to-noise ratios.

\section*{Acknowledgement}
We are grateful to Youichi Ohyama for valuable advice on the data reduction. We wish to thank Chris Pearson for much advice and for comments that improved the paper. We also thank the anonymous referee for many insightful comments. This work is based on observations obtained with the Apache Point Observatory 3.5-m telescope, which is owned and operated by the Astrophysical Research Consortium. This research was partially supported by the Japan Society for the Promotion of Science through Grant-in-Aids for Scientific Research (nos 16204013 and 18840047) and by a Sasakawa Scientific Research Grant from the Japan Science Society.

\label{lastpage}

\end{document}